\documentclass{emulateapj}
\usepackage{color}

\shorttitle{Pulsation Modes of Long-Period Variables}
\shortauthors{Soszy{\'n}ski, Wood \& Udalski}

\begin{document}

\title{Pulsation Modes of Long-Period Variables in the Period--Luminosity Plane}

\author{I. Soszy{\'n}ski\altaffilmark{1}, P. R. Wood\altaffilmark{2}, and A. Udalski\altaffilmark{1}}
\affil{$^1$Warsaw University Observatory, Al.~Ujazdowskie~4, 00-478~Warszawa, Poland\\soszynsk@astrouw.edu.pl, udalski@astrouw.edu.pl}
%\email{soszynsk@astrouw.edu.pl, udalski@astrouw.edu.pl}
\affil{$^2$Research School of Astronomy and Astrophysics, Australian National University, Cotter Road, Weston Creek ACT 2611, Australia\\wood@mso.anu.edu.au}
%\email{wood@mso.anu.edu.au}

\begin{abstract}
We present a phenomenological analysis of long-period variables (LPVs) in
the Large Magellanic Cloud with the aim of detecting pulsation modes
associated to different period--luminosity (PL) relations. Among brighter
LPVs we discover a group of triple-mode semiregular variables with the
fundamental, first overtone and second overtone modes simultaneously
excited, which fall on PL sequences C, C$'$ and B, respectively. The mode
identification in the fainter red giants is more complicated. We
demonstrate that the fundamental-mode pulsators partly overlap with the
first-overtone modes. We show a possible range of fundamental mode and
first overtone periods in the PL diagram.
\end{abstract}


\section{Introduction}

Stars that evolve on the first-ascent red giant branch (RGB) and asymptotic
giant branch (AGB) show, with no exception, variability due to
oscillations. The amplitudes of the photometric variations increase with
the bolometric luminosity of giant stars -- from micromagnitudes in stars
at the base of the giant branch \citep{hekker2009,bedding2010}, through
millimagnitudes in OGLE small amplitude red giants
\citep[OSARGs;][]{wray2004,soszynski2004}, tenths of a magnitude in
semiregular variables (SRVs) to several magnitudes in Mira stars at the tip
of the AGB. Recently, \citet{banyai2013} have shown that it is possible to
distinguish between solar-like oscillations low on the RGB and larger
amplitude pulsations which are characteristic of the more luminous OSARGs,
SRVs and Miras i.e. the OSARGs, SRVs and Miras all seem to be self-excited
pulsators rather than solar-like oscillators.

It is important to emphasize that there is no break in the overall red
giant evolutionary sequence, so one can expect to see a continuity between
the OSARGs, SRVs and Miras (collectively known as long-period variables --
LPVs). In the past, the OSARGs have sometimes been considered as a
separate class of variables from the SRVs and Miras
\citep[e.g.][]{soszynski2004,soszynski2007}.

Such continuity is clearly visible between the classically defined SRVs,
especially SRa stars, and Miras. Both groups follow the same
period--luminosity (PL) sequence, labeled C by \citet{wood1999}. The
difference between SRa stars and Miras lays in their amplitudes of
variability and in the number of excited pulsation modes. Miras are usually
single-mode pulsators, probably oscillating in the fundamental mode, while
SRa stars usually exhibit two modes -- fundamental and first overtone -- so
they occupy two sequences in the PL diagram, labeled C and C$'$.

However, LPVs display much more complex pattern in the PL plane. Recently,
\citet{soszynski2013} showed that SRb stars (smaller-amplitude and less
regular that SRa stars) can be found in the space between PL sequences C and
C$'$. A theoretical analysis suggests that SRb stars do not
follow a separate PL sequence, but they form a continuity with SRa stars
and Miras on sequence C. All of these stars are thought to pulsate in the
fundamental mode.

OSARG variables form a series of PL sequences \citep[e.g.][]{soszynski2007}
of which the longest-period ridge roughly overlaps with sequence C$'$. Even
so, it is not clear which modes of pulsation correspond to different PL
relations of OSARGs. For example, \citet{dziembowski2010} in their
theoretical work assumed that the longest-period PL sequence of OSARGs
corresponds to the fundamental mode. On the other hand,
\citet{takayama2013} argued that the same PL relation is associated with
the first-overtone. From the phenomenological point of view, the
identification of pulsation modes would be possible, if one detects LPVs
with the fundamental-mode (a period lying on sequence C), first-overtone
(sequence C$'$) and higher radial modes simultaneously excited in one
star. Up to this time, such a configuration has not been observed in the
pulsating red giants. LPVs with periods identified on sequence C have been
either single-mode (Miras) or double-mode (SRVs) pulsators.

In this paper we report the discovery of a small group of triple-mode LPVs
with the fundamental-mode, first-overtone and second-overtone
simultaneously excited. Our sample populates exclusively the brighter part
of the PL diagram and unambiguously assigns the pulsation modes to the PL
relations in this region. For fainter LPVs we found no distinct
counterparts of these triple-mode variables. We argue that sequence C$'$ in
its fainter part is populated by both -- fundamental mode and
first-overtone -- variables.

\vspace{3mm}
\section{A sample of LPVs in the Large Magellanic Cloud}

In our analysis, we use a huge collection of LPVs detected by the OGLE
project in the Large Magellanic Cloud \citep[LMC;][]{soszynski2009}. The
stars were observed between 2001 June and 2009 May with the 1.3~m Warsaw
telescope at the Las Campanas Observatory in Chile. The telescope was
equipped with the eight chip mosaic camera \citep{udalski2003}, covering
approximately $35'\times35'$ on the sky with the scale of
0.26~arcsec/pixel. Most of the images were obtained in the {\it I} band
with an exposure time of 180 seconds. The light curves consist typically of
700 points.

\begin{figure}
\epsscale{1.18}
\plotone{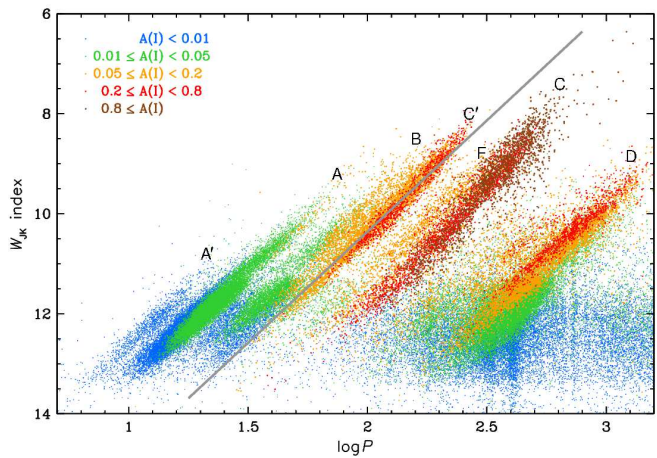}
\caption{Period--luminosity diagram for LPVs in the LMC. Each star is
  represented by one point, corresponding to the primary period. Different
  colors refer to different amplitudes: blue points show LPV with
  $A(I)<0.01$~mag, green points: $0.01\leq{A(I)}<0.05$~mag, orange points:
  $0.05\leq{A(I)}<0.2$~mag, red points: $0.2\leq{A(I)}<0.8$~mag, and brown
  points indicate Mira stars defined as LPVs with $A(I)\geq0.8$~mag. The grey
  solid line show the fit to PL sequence C$'$.}
\label{fig1}
\end{figure}

The photometry of LPVs in the central region of the LMC was supplemented
with the OGLE-II measurements collected between Jan 1997 and Nov 2000. We
also added observations obtained in the course of the ongoing OGLE-IV
project from 2010 March to 2013 May. Thus, the time baseline of
observations in some stars exceeded 16 years, and typically it was 12
years, with over 1000 observing points.

Our sample of LPVs was cross-matched with the 2MASS Point Source Catalog
\citep{cutri2003}. We left on our list only those stars that had both
$J$ and $K_s$-band measurements. For each object we derived the
reddening-independent near-infrared Wesenheit index, defined as
$$W_{JK}=K-0.686(J-K)~.$$

The distribution of LPVs in the period--$W_{JK}$ and period--$K$
diagrams look essentially the same for most SRVs and OSARGs. The main
difference between both diagrams concerns Miras with heavy circumstellar
extinction, which dominate the long-period end of sequence~C. In the
period--$K$ plane sequence~C is significantly broadened by the Miras
which are obscured by their circumstellar dust shells and lay far below
the linear PL relation \citep[e.g.][]{ita2011}. This effect may be
practically cancelled by using the reddening-independent Wesenheit
index, $W_{JK}$. Also the oxygen-rich and carbon-rich giants follow
nearly the same PL relations in the period--$W_{JK}$ plane
\citep{soszynski2007}, so our conclusions are valid for both spectral
types.

All the {\it I}-band light curves were searched for periodicities using the
Fourier-based Fnpeaks code by Z.~Ko{\l}aczkowski (private
communication). For each star we detected five periods with an iterative
procedure of fitting a third-order Fourier series and subtracting this
function from the light curve. For each period we recorded also the
amplitude of light variations, defined as a difference between maximum and
minimum value of the third-order Fourier series fitted to the observed
light curve.

The OGLE-III Catalog of LPVs in the LMC \citep{soszynski2009} counts
$91\;995$ objects, of which $79\;200$ were classified as (OSARGs),
$11\;128$ as SRVs and 1667 as Mira stars. Miras were distinguished from
SRVs on the basis of their {\it I}-band amplitudes larger than 0.8~mag. In
this study we do not separate OSARGs and SRVs, since we are trying to find
the continuity between both groups. We also had no effective method to
completely separate RGB and AGB OSARGs fainter than the tip of the RGB, so
we kept both these populations on the list of objects. RGB LPVs obey PL
relations which are somewhat shifted (in $\log{P}$) relative to the AGB
giants \citep{kiss2003}. However this offset does not change the
conclusions of this work, and in particular the identification of the
pulsation modes is valid for OSARGs on both the RGB and AGB.

\begin{figure}
\epsscale{1.19}
\plotone{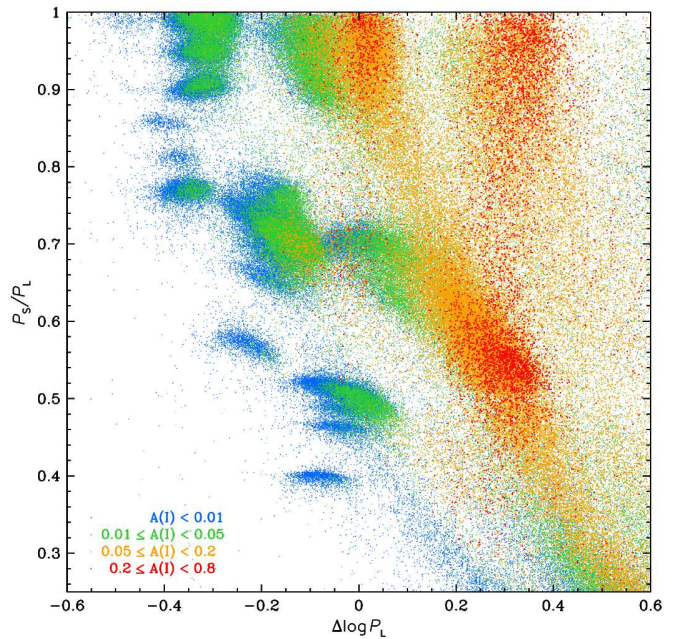}
\caption{Modified Petersen diagram for LPVs in the LMC.
  $P_{\mathrm S}/P_{\mathrm L}$ is the ratio of shorter to longer periods
  selected from up to five periods determined for each
  star. $\Delta\log{P_{\mathrm L}}$ indicates the horizontal distance of the
  longer period from the grey line plotted in the PL diagram
  (Fig.~\ref{fig1}).}
\label{fig2}
\end{figure}

\vspace{3mm}
\section{Analysis}

Fig.~\ref{fig1} shows the period--Wesenheit index diagram for LPVs in the
LMC. Each star is represented by only one point corresponding to the
primary period (first period detected by the Fnpeaks code -- usually
associated with the largest amplitude). Different colors of the points code
different ranges of {\it I}-band amplitude -- from OSARGs with
$A(I)<0.01$~mag represented by blue symbols to Miras ($A(I)>0.8$~mag)
indicated by brown points. Sometimes the automatically detected periods
are spurious, which is particularly evident for some of the
smallest-amplitude variables (blue points), which show pseudo-periods
longer than sequence D. However, most of the shorter periods are real and
trace in Fig.~\ref{fig1} the clearly visible PL sequences, which are
labeled according to the scheme introduced by \citet{wood1999} and other
authors. Mira stars follow sequence C, SRVs -- sequences C and C$'$, OSARGs
-- sequences A, A$'$, and B. The grey line overplotted in Fig.~\ref{fig1}
has a slope $dW_{JK}/d\log{P}=-4.444$ and it is a rough fit to sequence
C$'$.

With the exception of Mira stars, LPVs are known to be multi-periodic
variable stars. A widely used tool for studying multi-mode variables is the
so-called Petersen diagram, on which the ratio of two selected periods is
plotted against the logarithm of the longer period. However, LPVs form a
quite fuzzy picture on the classical Petersen diagram, because periods and
period ratios corresponding to different PL sequences overlap each other.

To circumvent this problem, we plotted a somewhat different diagram
(Fig.~\ref{fig2}) which we called a modified Petersen diagram. In this
diagram the ordinate axis shows (as in the classical Petersen diagram) the
ratio of shorter to longer periods $P_{\mathrm S}/P_{\mathrm L}$ selected
from the five most significant periods determined in every star, while the
abscissa axis shows the horizontal distance (in $\log{P_{\mathrm L}}$) of a
star from the grey solid line plotted in Fig.~\ref{fig1}. In the modified
Petersen diagram different pairs of periods corresponding to different PL
relations form distinct groups of points. In Fig.~\ref{fig2} colors of the
points indicate amplitudes associated with the longer periods. Since Mira
stars are usually single-mode pulsators, they are not included in the
modified Petersen diagram.

\begin{figure}
\epsscale{1.17}
\plotone{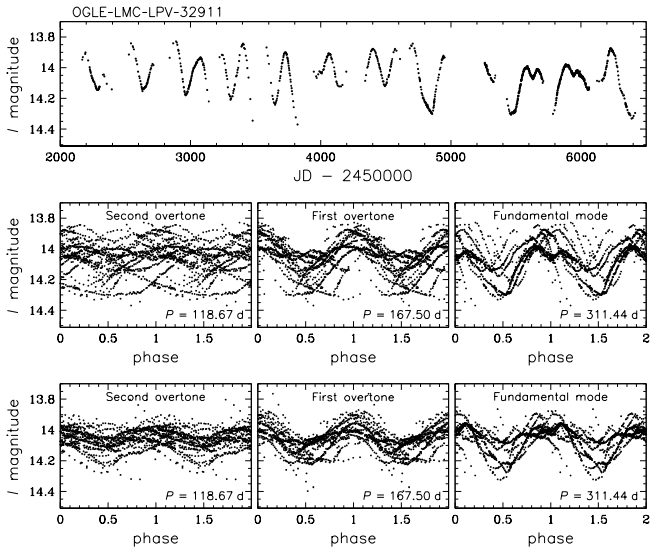}
\caption{Light curve of a triple-mode SRV OGLE-LMC-LPV-32911. Upper panel
  shows an unfolded light curve, middle panels present the same light curve
  folded with tree pulsation periods, and lower panels show the light
  curves pre-whitened with the four other detected periods. Periods and
  mode identifications are given in the panels.}
\label{fig3}
\end{figure}

\vspace{3mm}
\section{Pulsation modes of LPVs}

Assignment of the pulsation modes to different PL relations formed by LPVs
is a matter of debate
\citep[e.g.][]{wood1996,wood1999,kiss2003,ita2004,takayama2013}. There is a
consensus today that Mira stars pulsate in the fundamental mode. Miras are
generally single-mode variables and lie on sequence C in the PL plane, so
other LPVs with one of their periods falling on this sequence are also
considered as fundamental-mode pulsators. Most of SRVs are double-mode
variables with the fundamental mode and first overtone simultaneously
excited. The first-overtone periods of SRVs delineate sequence C$'$ in the
PL diagram. In the modified Petersen diagram (Fig.~\ref{fig2}) SRV of type
SRa (red points) concentrate at period ratios between 0.45 and 0.6. SRb
stars with smaller amplitudes (orange points) exhibit on average larger
ratios of the first-overtone to fundamental-mode periods. Their
fundamental-mode periods can be found somewhere between sequences C and
C$'$. \citet{soszynski2013} labeled this fundamental-mode sequence of SRb
stars by letter F, although it seems that sequences C and F constitute a
continuity between each other.

The identification of the pulsation modes associated with the
shorter-period PL relations obeyed by LPVs remains a subject of
controversy. A definite answer would be given by multi-mode LPVs with the
fundamental, first-overtone and higher overtone simultaneously excited.
However no such objects have been reported so far. LPVs with one of their
periods lying on sequence C (fundamental-mode) have been found to be either
single-mode Miras or double-mode SRVs.

\begin{figure}
\epsscale{1.17}
\plotone{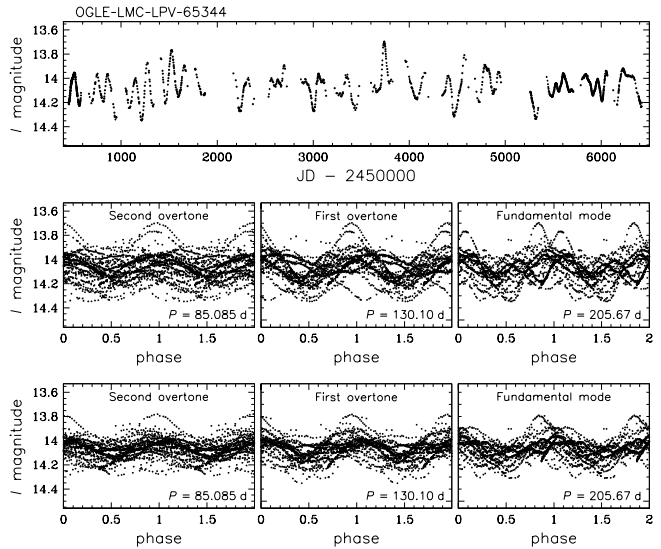}
\caption{Light curve of a triple-mode SRV OGLE-LMC-LPV-65344. Upper panel
  shows an unfolded light curve, middle panels present the same light curve
  folded with tree pulsation periods, and lower panels show the light
  curves pre-whitened with the four other detected periods. Periods and
  mode identifications are given in the panels.}
\label{fig4}
\end{figure}

In this regard, we performed a search for SRVs with at least three radial
modes simultaneously excited. We divided the PL plane into three regions:
one covering the fundamental-mode periods (sequences C and F), the second
comprising the first-overtone periods (sequence C$'$), and the third region
covering the shorter-period sequences (A$'$, A and B). As mentioned above,
we derived five periods for each star from our sample of LPVs in the
LMC. From this sample we selected those objects which had at least one
period in each of these regions. From our analysis we excluded periods
longer than those making up sequence C: still unexplained long secondary
periods which comprise sequence D, periods which populate a newly
discovered dim sequence located between sequences C and D
\citep{soszynski2013}, and sequence E consisting of binary systems.

Since some of the automatically derived periods are spurious, we visually
inspected all the selected light curves. For each period we pre-whitened
the light curves with four other detected periods and subjectively decided
whenever the remaining period is real. Two examples of such light curves
are shown in Figs.~\ref{fig3} and~\ref{fig4}. In the upper panel of these
figures we show unfolded light curves, middle panels show original light
curves folded with three periods corresponding to three pulsation modes,
and bottom panels present pre-whitened light curves folded with the same
periods.

\begin{figure}
\epsscale{1.19}
\plotone{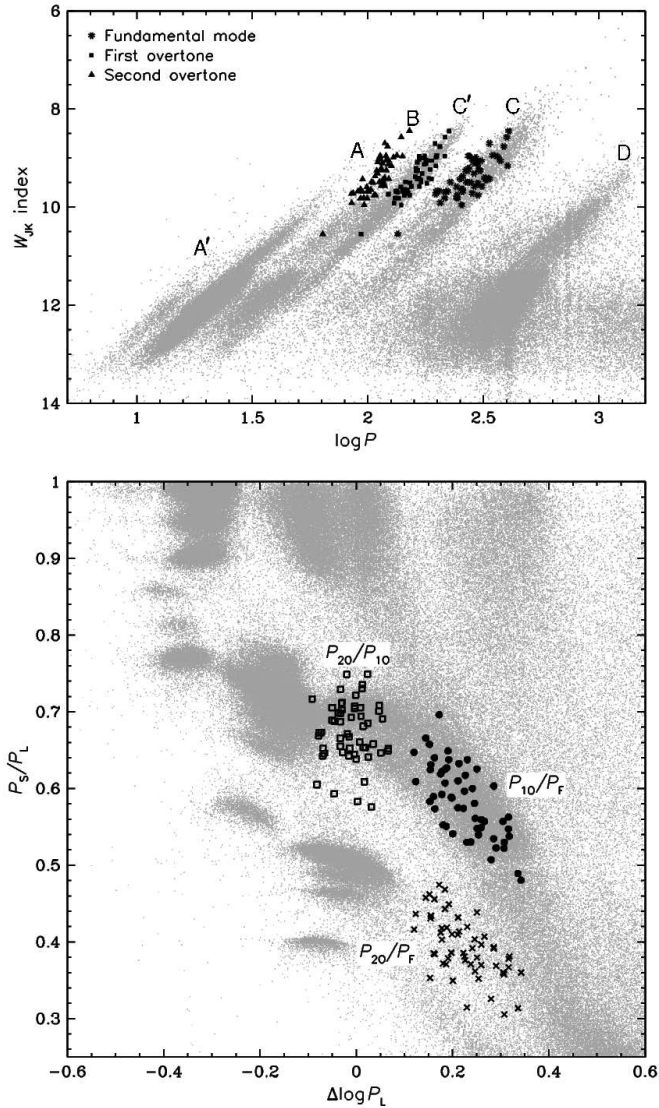}
\caption{Period--luminosity diagrams (upper panel) and modified Petersen
  diagram (lower panel) for triple-mode SRVs with the fundamental-, first-
  and second-overtone modes excited. Different symbols indicate different
  periods (upper panel) and period ratios (lower panel). Grey points in the
  background present all LPVs in the LMC.}
\label{fig5}
\end{figure}

As a result of the visual examination we detected a limited number of about
50 candidates for triple-mode LPVs -- with fundamental mode, first and
second overtones excited. The location of these stars in the PL diagram and
modified Petersen diagram is shown in Fig.~\ref{fig5}. The fundamental-,
first- and second-overtone modes fall on sequences C, C$'$ and B,
respectively. The most important aspect of this result is that it
unambiguously links together the SRVs and Miras from sequences C and C$'$
and the OSARG variables from sequence B (as well as A and A$'$).

It is striking that in spite of the fact that we
analysed the whole sample of LPVs, we identified reliable candidates for
triple-mode pulsators exclusively among brighter stars. Only one star in
our sample is fainter than $W_{JK}=10$~mag. In the lower part of the PL
diagram we found a number of LPVs with the longest period falling on
sequence C and two closely-spaced periods located roughly on sequence
C$'$. Such a closely-spaced periods were also detected in the OSARG variables
populating sequences A and B \citep{soszynski2004,soszynski2007}. Such a
phenomenon may be caused by non-radial pulsation modes or by small
variations of the pulsation periods, but not by consecutive radial modes.

Thus, in the lower part of the PL diagram we found no reliable examples of
triple-mode pulsators, at least with the longest period falling on sequence
C or F. It complicates the identification of the pulsation modes in this
region. Most of SRVs have two periods: one falling on sequence C and the
second lying on sequence C$'$.

\begin{figure}
\epsscale{1.19}
\plotone{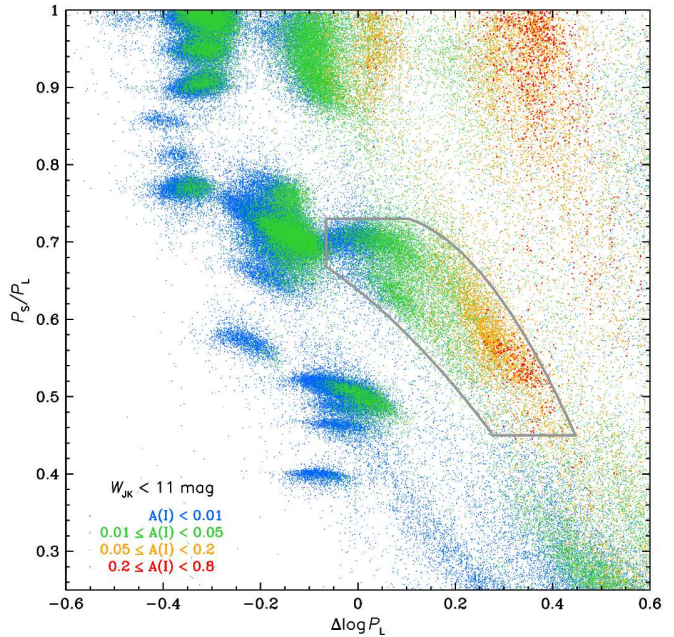}
\caption{Modified Petersen diagram for LPVs fainter than
  $W_{JK}=11$~mag. Different colors refer to different amplitudes, as in
  Fig.~2. The grey contour indicates the area where $P_{1O}/P_F$ and
  $P_{2O}/P_{1O}$ ratios are expected.}
\label{fig6}
\end{figure}

Fig.~\ref{fig6} displays the modified Petersen diagram for LPVs fainter
than $W_{JK}=11$~mag. The grey contour in this diagram roughly surrounds
the area where $P_F/P_{1O}$ and $P_{2O}/P_{1O}$ ratios were observed for
brighter SRVs. As can be seen, there is no natural boundary that separates
both groups, fundamental-mode/first-overtone and
first-overtone/second-overtone pulsators overlap in the modified Petersen
diagram. It means that the PL sequence associated with the fundamental mode
extends much further toward the shorter periods than shown in the paper by
\citet{soszynski2013}. Probably some of the fundamental-mode periods fall
on sequence C$'$ which had previously been associated with the
first-overtone mode.

So, where is the short-period limit of the fundamental pulsation mode of
LPVs? The definitive answer cannot be given just on the basis of the
photometric observations, because it seems that different modes of
pulsations of LPVs may overlap in the PL, Petersen and other empirical
diagrams. This is because in the LMC we observe a mixture of red giants
with different masses and in different evolutionary stages.

\begin{figure}
\epsscale{1.17}
\plotone{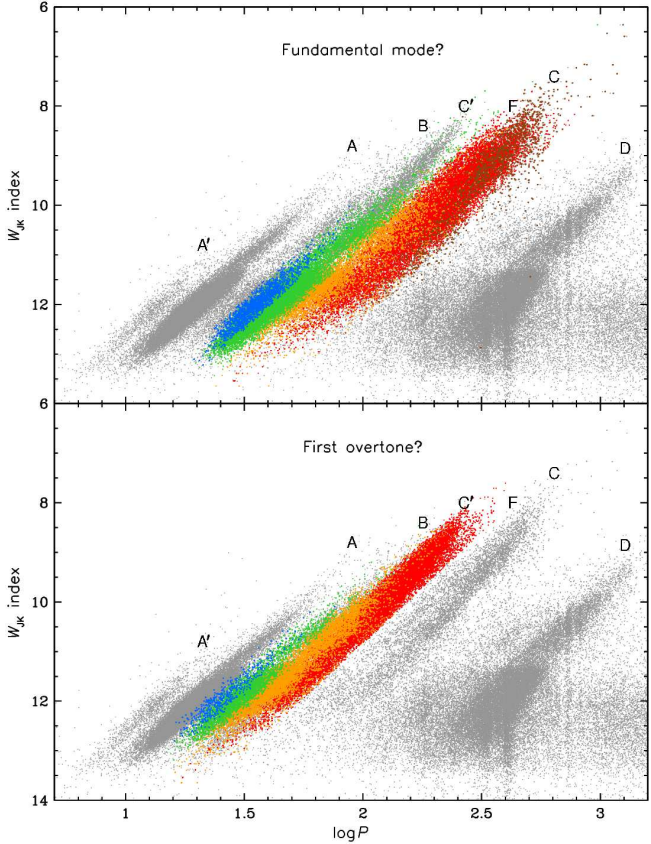}
\caption{Period--luminosity diagrams for the potential fundamental-mode
  (upper panel) and first-overtone (lower panel) periods of LPVs in the
  LMC. The points in the upper panel are the longest pulsation periods
  detected in each star while the points in the lower panel are second
  longest periods. Brown points indicate Mira stars ($A(I)>0.8$~mag), red
  points -- SRVs, orange points -- OSARGs without a period on sequence A or
  sequence A$'$, green points -- OSARGs with one of their periods falling
  on sequence A, but not on sequence A$'$, and blue points -- OSARG with at
  least one period lying on sequence A$'$. Grey points in the background
  present the primary periods of all LPVs from our sample.}
\label{fig7}
\end{figure}

A clue about the range of the fundamental mode (and other modes) in the PL
diagram can be given by the analysis of the longest pulsation periods of
various LPVs (upper panel of Fig.~\ref{fig7}). As previously noted,
fundamental-mode periods of SRVs and Miras populate sequence C and space
between sequences C and C$'$ (red and brown points in
Fig.~\ref{fig7}). OSARG variables obey a series of PL relations, of which
the longest-period sequence roughly coincides with sequence C$'$ (ignoring
sequence D which is not associated with any radial pulsation
mode). However, the exact position of this PL relation depends on other
pulsation periods of these multi-periodic variables. If we select OSARGs
that do not have a period falling on either sequence A or sequence A$'$,
and we plot them in the PL diagram, we can see two ridges, of which the
longer one is located slightly to the right from sequence C$'$ (orange
points in Fig.~\ref{fig7}). OSARG stars which have a detectable period on
sequence A, but do not have a detectable period on sequence A$'$, have
their longest period more or less at the location of sequence C$'$ (green
points in Fig.~\ref{fig7}). Finally, OSARG variables with at least one of
their periods on sequence A$'$ follow four PL relations, of which the
longest one is located to the left from sequence C$'$ (blue points in
Fig.~\ref{fig7}). Although we are not sure that the longest period in all
these LPVs correspond to the fundamental mode, this continuity in the
position of the longest-period sequences suggests that the longest period
is the fundamental mode.

In the lower panel of Fig.~\ref{fig7} we plot the PL sequences of the same
stars but using the second longest pulsation period (with exception of
Miras which are single-mode variables). These are candidates for
first-overtone mode pulsation. As one can see, the fundamental mode and
first overtone LPVs significantly overlap in the PL diagram.

\vspace{3mm}
\section{Conclusions}

In this paper we tried to assign pulsation modes to LPVs on different PL
relations. For brighter SRVs we showed that sequences C, C$'$, and B are
populated by the fundamental mode, first overtone, and second overtone,
respectively. For fainter LPVs the fundamental modes of some LPVs overlap
with the first-overtones of other pulsating red giants. The region of the
PL plane occupied by the longest period of different types of LPVs forms a
continuous wide ridge which spreads from OSARG variables (sequence C$'$),
through SRb stars \citep[sequence F of][]{soszynski2013} to SRa and
Mira stars (sequence C). It is likely that all these periods are caused by
the fundamental-mode pulsations, which is in agreement with the theoretical
investigation of \citet{dziembowski2010} and is opposite to the mode
identifications by \citet{takayama2013}. Similarly, the first-overtone
pulsators (second longest periods) fall in a region that covers sequences
B and C$'$. The exact position of different modes in the PL plane depends
on the mass, metallicity and evolutionary status of a star. Variations in
the mass and metallicity are the likely cause of the broad PL sequences.

One of the most important result of this paper is showing that OSARGs, SRVs
and Miras constitute a continuity, with no break between these types of
pulsating red giants. As a giant star evolves, it gradually changes from
OSARG to SRV and finally to Mira, increasing its pulsating periods,
limiting the number of excited modes, and increasing its amplitudes of
variability. It seems that the boundary between OSARG variables and SRVs is
arbitrary, as it is between SRVs and Miras (where the boundary is defined
as the amplitude in the V band equal to 2.5~mag).

\vspace{3mm}
\section*{Acknowledgments}

IS is indebted to Wojciech A. Dziembowski for many fruitful discussions on
the topic of this paper. This work has been supported by the Polish
Ministry of Science and Higher Education through the program ``Ideas Plus''
award No. IdP2012 000162. The research leading to these results has
received funding from the European Research Council under the European
Community's Seventh Framework Programme (FP7/2007-2013)/ERC grant agreement
no. 246678.

\label{lastpage}

\end{document}